\documentclass[10pt]{article}

\usepackage{amstext,amsmath,amssymb,amsfonts,bbm,slashed}
\usepackage[latin1]{inputenc}
\usepackage{epsfig}
\usepackage{hyperref}
\usepackage{amsthm}
\usepackage{subfigure}
\usepackage{color}
\usepackage{multirow}

\newtheorem{lemma}{Lemma}

\newtheorem{theorem}{Theorem}
\newtheorem{proposition}{Proposition}

\newcommand{\bea}{\begin{eqnarray}}
\newcommand{\eea}{\end{eqnarray}}
\newcommand{\beq}{\begin{equation}}
\newcommand{\eeq}{\end{equation}}

\newcommand{\bpro}{\begin{pro}}
	\newcommand{\epro}{\end{pro}}
\newcommand{\blem}{\begin{lem}}
	\newcommand{\elem}{\end{lem}}
\newcommand{\bdfn}{\begin{dfn}}
	\newcommand{\edfn}{\end{dfn}}
\newcommand{\bcor}{\begin{cor}}
	\newcommand{\ecor}{\end{cor}}
\newcommand{\bthm}{\begin{thm}}
	\newcommand{\ethm}{\end{thm}}
\newcommand{\bex}{\begin{ex}}
	\newcommand{\eex}{\end{ex}}
\newcommand{\brmk}{\begin{rmk}}
	\newcommand{\ermk}{\end{rmk}}
\newcommand{\bpr}{\begin{pr}}
	\newcommand{\epr}{\end{pr}}

\makeatletter

\begin{document}

	\begin{center}
		
		{\LARGE\bf  Recursion operator in a noncommutative\\  Minkowski phase space}

		\vspace{15pt}
		
		{\large   Mahouton Norbert Hounkonnou, Mahougnon Justin Landalidji and Ezinvi Balo\"{\i}tcha
		}
		
		\vspace{15pt}
		
		{ 
			International Chair of Mathematical Physics
			and Applications  (ICMPA-UNESCO Chair)\\
			University of Abomey-Calavi, 072 B.P. 50 Cotonou, Republic of Benin\\
			E-mails: { norbert.hounkonnou@cipma.uac.bj with copy to
				hounkonnou@yahoo.fr} \\
			E-mails: { landalidjijustin@yahoo.fr}   \\
			E-mails: {       ezinvi.baloitcha@cipma.uac.bj} 
		}

	\end{center}
	\vspace{10pt}
	\begin{abstract}
		A recursion operator for a geodesic flow, in a noncommutative (NC) phase space endowed with a Minkowski metric, is constructed and discussed in this work.
		A NC Hamiltonian function $\mathcal{H}_{nc}$  describing the dynamics of a free particle system in such a phase space, equipped  with a noncommutative symplectic form $\omega_{nc}$ is defined. A related NC Poisson bracket is obtained. This permits to construct the NC Hamiltonian vector field, also called NC geodesic flow. Further, using a canonical transformation induced by a generating function from the  Hamilton-Jacobi equation, we obtain a relationship between  old and  new coordinates, and their conjugate momenta. These new coordinates  are used to re-write the NC recursion operator in a simpler form, and  to deduce the corresponding constants of motion. Finally, all obtained physical quantities are re-expressed and analyzed in the initial NC canonical coordinates.
		
		
	\end{abstract}
	
	\section{Introduction}\label{sec_1}
	In the last few decades there was a renewed interest in completely integrable
	Hamiltonian systems, the concept of which goes back to 
	Liouville in 1897 \cite{li} and  Poincar\'{e} in 1899 \cite{po} . They are dynamical systems admitting a Hamiltonian description and possessing sufficiently many constants of motion, so that they can be integrated by quadratures. Some qualitative features of these systems remain true in some special classes of infinite-dimensional Hamiltonian systems expressed by nonlinear evolution equations as, for instance,
	Korteweg-de Vries and Sine-Gordon \cite{vil1} . 
	
	A relevant progress in the study of these systems with an infinite-dimensional
	phase manifold $\mathcal{M}$ was the introduction of the Lax Representation \cite{lax}.
	It played an important role in formulating the Inverse Scattering Method \cite{ablo}, one of the most remarkable result of theoretical physics in last decades. This method allows
	the integration of nonlinear dynamics, both with  finitely or infinitely many degrees of freedom, for which a Lax representation can be given \cite{ga}, this being both of physical and mathematical relevance \cite{fa}. 
	
	Another progress, in the analysis of the integrability, was the important
	remark that many of these systems are Hamiltonian dynamics with respect to two compatible symplectic structures  \cite{mag}, \cite{gel}, \cite{vil1}, this leading to a geometrical interpretation of the so-called recursion operator \cite{lax}. For more details, see \cite{spa} and references therein. A description of integrability
	working both for systems with finitely many degrees of freedom and for field theory
	can be given in terms of invariant, diagonalizable mixed $(1,1)$-tensor field, having bidimensional eigenspaces and vanishing Nijenhuis torsion.
	A natural approach to integrability is to try to find sufficient conditions for the eigenvalues of the recursion operator to be in involution. Thereby, a new characterization of integrable Hamiltonian systems is given by De Filippo {\it et al} through  the following Theorem \cite{fil1}:
	\begin{theorem}\label{thi}
		Let $X$ be a dynamical vector field on a $2n$-dimensional manifold $\mathcal{M}$. If the vector field $X$ admits a diagonalizable mixed $(1, 1)$-tensor field $T$ which is invariant under $X,$ has a vanishing Nijenhuis torsion and has doubly degenerate eigenvalues
		with nowhere vanishing differentials, then there exist a symplectic structure and a
		Hamiltonian function $H$ such that the vector field $X$ is separable, Hamiltonian vector
		field of $H,$ and $H$ is completely integrable with respect to the symplectic structure.
	\end{theorem}
	Such a $(1, 1)$-tensor field $T$ is called a recursion operator of $X$.
	\paragraph{}In a particular case of $\mathbb{R}^{2n},$ a recursion operator can be constructed as follows \cite{ta}:
	\begin{lemma} \label{lem_1} \
		Let us consider vector fields
		\[
		X_{l} = - \dfrac{\partial}{\partial{x_{n+l}}}, \ \ l = 1,...,n
		\]
		on $\mathbb{R}^{2n}$ and let $T$ be a $(1,1)$-tensor field on $\mathbb{R}^{2n}$ given by
		\[
		T = \sum_{i =1}^{n}x_{i}\Bigg(\dfrac{\partial}{\partial{x_{i}}} \otimes dx_{i} + \dfrac{\partial}{\partial{x_{n+i}}} \otimes dx_{n+i}\Bigg).
		\]
		Then, we have that the Nijenhuis torsion $ \mathcal{N}_{T}$ and the Lie derivative $\mathcal{L}_{X_{l}}$ of {$T$} are vanishing,  i. e. 
		\begin{eqnarray} (\mathcal{N}_{T})^{h}_{ij}:=T^{k}_{i}\dfrac{\partial{T^{h}_{j}}}{\partial{x^{k}}}  - T^{k}_{j}\dfrac{\partial{T^{h}_{i}}}{\partial{x^{k}}} + T^{h}_{k}\dfrac{\partial{T^{k}_{i}}}{\partial{x^{j}}}  - T^{h}_{k}\dfrac{\partial{T^{k}_{j}}}{\partial{x^{i}}} &=& 0,\cr
		\mathcal{L}_{X_{l}}T& =& 0.
		\end{eqnarray}	
		That is  the $(1,1)$-tensor field $T$ is a recursion operator of $X_{l}, (l = 1,...,n).$
	\end{lemma}
	
	On the other hand, this $(1, 1)$-tensor field $T$ is used as an operator which generate enough constants of motion \cite{gl}. Based on Theorem \ref{thi} , a series of  investigations was done (see e.g \cite{fil1}, \cite{fil2}, \cite{gl},  \cite{mar}, \cite{spa}, \cite{vil2}, \cite{kh}, \cite{t},\cite{ta}, \cite{ta2} and references therein). One of powerful methods of describing completely  integrable Hamiltonian systems with involutive Hamiltonian functions or constants of motion uses the recursion operator admitting a vanishing Nijenhuis torsion.
	
	Recently, in 2015,  Takeuchi  constructed  recursion operators of Hamiltonian vector fields of geodesic flows for some  Riemannian and Minkowski  metrics \cite{t}, and  obtained related constants of motion.
	Further, he   used five particular solutions of the Einstein equation in the Schwarzschild,  Reissner-Nordström,   Kerr,  Kerr-Newman, and  FLRW metrics, and showed that the Hamiltonian
	functions of the associated corresponding geodesic flows  form a system of variables separation  equations. Then, he  constructed recursion operators  inducing the complete integrability of the Hamiltonian functions.
	In the present work, we investigate  the same problem in a deformed Minkowski phase space.
	
	This paper is organized as follows.  In Section \ref{sec_2}, we consider  a noncommutative Minkowski phase space,  and define the NC Hamiltonian function and symplectic form, as well as the corresponding NC Poisson bracket. In Section \ref{sec_3},  we construct the associated NC recursion operator for the NC Hamiltonian vector field of the geodesic flow, and obtain related constants of motion. In Section \ref{sec_4}, we end with  some concluding remarks.
	
	\section{Noncommutative Minkowski phase space} \label{sec_2}
	
	The noncommutativity between space-time coordinates was first introduced by Snyder \cite{sy}. Later, Alain Connes  developed the noncommutative geometry \cite{alinc1} and applied it to various physical situations \cite{alinc2}. Since then, the noncommutative geometry remained  a very active research subject in several domains of theoretical physics and mathematics.
	\paragraph{}Noncommutativity between phase space variables is here understood by replacing the usual product with the $\beta-$star product, also known as the Moyal product law between two arbitrary functions of position and momentum, as follows
	\cite{vak}, \cite{ma}, \cite{khos1} :
	\begin{equation} \label{Eq_3_1}
	(f\ast_{\beta}g) (q,p) = f(q_{i},p_{i}) \exp{\bigg(\dfrac{1}{2}\beta^{ab}\overleftarrow{\partial}_{a}\overrightarrow{\partial_{b}}\bigg)} g(q_{j},p_{j})\Bigg|_{(q_{i},p_{i})= (q_{j},p_{j})},
	\end{equation}
	where
	\begin{equation} \label{Eq_3_2}
	\beta_{ab} = \left(
	\begin{array}{cc}
	\alpha_{ij} & \delta_{ij} + \gamma_{ij} \\
	- \delta_{ij} - \gamma_{ij} & \lambda_{ij} \\
	\end{array}
	\right),
	\end{equation}
	$\alpha$ and $\lambda$ are antisymmetric  $n \times n$ matrices which represent the noncommutativity in coordinates and momenta, respectively;   $\gamma$ is some  combination of $\alpha$ and $\lambda.$ 
	The 	$\ast_{\beta}$ deformed Poisson bracket can be written as 
	\begin{equation} \label{Eq_3_3}
	\{f,g\}_{\beta} = f\ast_{\beta}g - g\ast_{\beta}f.
	\end{equation}
	So,  we can  show that :
	\begin{equation}\label{Eq_3_4}
	\{q_{i},q_{j}\}_{\beta} = \alpha_{ij}, \ \{q_{i},p_{j}\}_{\beta} =\delta_{ij} + \gamma_{ij}, \ \{p_{i},q_{j}\}_{\beta} = - \delta_{ij} - \gamma_{ij}, \ \{p_{i},p_{j}\}_{\beta} =\lambda_{ij}.
	\end{equation}
	Now, consider the following transformations :
	\begin{equation}\label{Eq_3_5}
	q'_{i} = q_{i} - \dfrac{1}{2}\sum_{j=1}^{n}\alpha_{ij}p_{j}, \quad p'_{i} = p_{i} + \dfrac{1}{2}\sum_{j=1}^{n}\lambda_{ij}q_{j},
	\end{equation}
	where 	$q'_{i}$ and 	$p'_{j}$ obey the same commutation relations as in  \eqref{Eq_3_4}, but with respect to the usual Poisson bracket :
	\begin{equation} \label{Eq_3_6}
	\{q'_{i},q'_{j}\} = \alpha_{ij}, \ \{q'_{i},p'_{j}\}=\delta_{ij} + \gamma_{ij}, \ \{p'_{i},q'_{j}\}= -\delta_{ij} - \gamma_{ij}, \ \{p'_{i},p'_{j}\} =\lambda_{ij},
	\end{equation}
	with $q_{i}$ and $p_{j}$ satisfying the following commutation relations :
	\begin{equation} \label{Eq_3_7}
	\{q_{i},q_{j}\} = 0, \quad \{q_{i},p_{j}\}=\delta_{ij}, \quad \{p_{i},p_{j}\}=0.
	\end{equation}
	\paragraph{}	In our framework, we consider the noncommutative Minkowski phase space with the metric  defined by
	\begin{equation}
	ds'^{2} = -dq_{1}'^{2} + dq_{2}'^{2} + dq_{3}'^{2} + dq_{4}'^{2},\end{equation} 
	where $q'_{1} \ \mbox{is time coordinate}; \ q'_{2},\ q'_{3},\ q'_{4} \  \mbox{are space coordinates}.$
	The tensor metric is given by
	\begin{equation}
	g'_{ij} = g'^{ij} = \left(
	\begin{array}{cccc}
	-1 & 0 & 0 & 0 \\
	0 & 1& 0 & 0 \\
	0 & 0 & 1 & 0 \\
	0 & 0 & 0 & 1 \\
	\end{array}
	\right),
	\end{equation}
	and the equation of geodesic is
	\begin{equation}
	\dfrac{d^{2}q'^{\mu}}{dt^{2}} + \Gamma'^{\mu}_{\nu\lambda}\dfrac{dq'^{\nu}}{dt}\dfrac{dq'^{\lambda}}{dt} = \dfrac{d^{2}q'^{\mu}}{dt^{2}}= 0, \ \ (\mu = 1,2,3,4),\end{equation}
	with 
	the Christoffel symbols  $\Gamma'^{\mu}_{\nu\lambda} = 0$.  \\
	Set:							 
	\begin{equation}\label{Eq_3_7_1}
	q'_{i} = q_{i} - \dfrac{1}{2}\sum_{j = 1}^{4}\alpha_{ij}p_{j}, \ \  p'_{i} = p_{i} +  \dfrac{1}{2}\sum_{j = 1}^{4}\lambda_{ij}q_{j},\;\lambda_{1j} = \alpha_{1j} = 0, \  p_{1}> 0.
	\end{equation}
	Then, the commutation relations 	\eqref{Eq_3_6} become :
	\begin{equation}\label{nc1}
	\{q'_{i},q'_{j}\} = \alpha_{ij}, \  \{q'_{i},p'_{j}\}=\delta_{ij} + \gamma_{ij},\  \{p'_{i},q'_{j}\}= -\delta_{ij} - \gamma_{ij}, \  \{p'_{i},p'_{j}\} =\lambda_{ij}.
	\end{equation}
	
	\subsection{NC Hamilton function and NC symplectic form}
	The  NC Hamiltonian function  $\mathcal{H}_{nc}$ describing the dynamics of a free particle system, in the considered NC Minkowski phase space is defined as follows :
	\[
	\mathcal{H}_{nc} :=\dfrac{1}{2}\Bigg(-p_{1}'^{2} + \sum_{k=2}^{4}p'^{2}_{k}\Bigg).
	\]
	
	Using equations \eqref{Eq_3_7_1}, we get 
	\begin{equation}\label{Eq_3_8}
	\mathcal{H}_{nc} =\dfrac{1}{2}\Bigg[ -p_{1}^{2} + \sum_{k=2}^{4} \Bigg(p_{k}+\dfrac{1}{2}\sum_{j = 2}^{4}\lambda_{kj}q_{j}\Bigg)^{2} \Bigg].
	\end{equation}
	\begin{proposition}
		The exterior derivative of the Hamiltonian function $ \mathcal{H}_{nc}$ is given by
		\begin{align} \label{dh}
		d\mathcal{H}_{nc}  = - p_{1}dp_{1} + \sum_{k =2}^{4}\varpi_{k}dp_{k} 
		+\dfrac{1}{2}\sum_{k,i =2}^{4}\lambda_{ik}\Omega_{i}dq_{k},
		\end{align}
		where 
		\[ \varpi_{k} = \Bigg(p_{k} + \dfrac{1}{2}\sum_{j=2}^{4}\lambda_{kj}q_{j}\Bigg)
		\]
		and
		\[  \Omega_{i} = \Bigg(p_{i} + \dfrac{1}{2}\sum_{j = 2}^{4}\lambda_{ij}q_{j}\Bigg). \] 
	\end{proposition}
	
	\paragraph{}The NC symplectic form is now defined by 
	\begin{equation}
	\omega_{nc} := \displaystyle\sum_{i=1}^{4}dp'_{i}\wedge dq'_{i} = dp'_{1}\wedge dq'_{1} + \displaystyle\sum_{k=2}^{4}dp'_{k}\wedge dq'_{k}.					\end{equation}
	\begin{proposition}
		Considering the NC Minkowski phase space, the symplectic form associated with the  Hamiltonian function $ \mathcal{H}_{nc}$ is given by
		\begin{equation}\label{Eq_3_9}	\omega_{nc} = \sum_{\nu=1}^{4}\theta_{\nu}dp_{\nu}\wedge dq_{\nu},		\end{equation}
		where 	
		\begin{equation}\label{sys2}	\theta_{\nu} = \displaystyle\sum_{\mu = 1}^{4}\Bigg(\delta_{\mu\nu} + \dfrac{1}{4}\lambda_{\mu\nu}\alpha_{\mu\nu}\Bigg) \neq 0, \ \ \delta_{\mu\nu} = \left\{
		\begin{array}{ll}
		0, & \mbox{if} \ \mu \neq \nu \\
		1, & \mbox{if} \ \mu = \nu.
		\end{array}
		\right.
		\end{equation}	
		
	\end{proposition}

	\subsection{NC Poisson bracket and NC Hamiltonian vector field}
	
	\begin{proposition}
		The bracket given by
		\begin{equation} \label{Eq_3_10}
		\{f,g\}_{nc} = \sum_{\nu=1}^{4}\theta_{\nu}^{-1}\Bigg(\dfrac{\partial{f}}{\partial{p_{\nu}}}\dfrac{\partial{g}}{\partial{q_{\nu}}} - \dfrac{\partial{f}}{\partial{q_{\nu}}}\dfrac{\partial{g}}{\partial{p_{\nu}}}\Bigg)
		\end{equation}
		is a Poisson bracket which respects the symplectic form $\omega_{nc},$ where $f$ and $g$ are arbitrary differentiable  coordinate functions on the NC Minkowski phase space.
	\end{proposition} 
	
	\begin{proposition}
		In the NC Minkowski phase space, the Hamiltonian vector field is  given by
		\begin{equation}\label{Eq_3_16}
		X_{\mathcal{H}_{nc}}  = - p_{1}\dfrac{\partial}{\partial{q_{1}}} + \sum_{k=2}^{4}\theta^{-1}_{k}\Bigg(\varpi_{k}\dfrac{\partial}{\partial{q_{k}}}  - \dfrac{1}{2}\sum_{i=2}^{^4}\lambda_{ik}\Omega_{i}\dfrac{\partial}{\partial{p_{k}}}\Bigg), 
		\end{equation}
		where
		\[
		\varpi_{k} = \bigg(p_{k} + \dfrac{1}{2}\sum_{j=2}^{4}\lambda_{k j}q_{j}\bigg)\]
		and
		\[
		\Omega_{i}= \bigg(p_{i} + \dfrac{1}{2}\sum_{j = 2}^{4}\lambda_{ij}q_{j}\bigg).
		\]
	\end{proposition}
	
	\section{NC Recursion operator} \label{sec_3}
	
	In this section, we construct the recursion operator for the geodesic flow in the NC Minkowski phase space, and derive the constants of motion. We consider the Hamilton-Jacobi equation \cite{arn} for the Hamiltonian function \eqref{Eq_3_8}, and  introduce a generating function $W^{nc}$ satisfying the following relations:
	\begin{equation}\label{can}
	p = \dfrac{\partial{W^{nc}}}{\partial{q}} \quad \mbox{and} \quad P = - \dfrac{\partial{W^{nc}}}{\partial{Q}}. 
	\end{equation}
	This allows us to obtain a relationship between the $(p,q)$ and  $(P,Q)$ coordinates.
	Then, using Lemma \ref{lem_1}, we build  a recursion operator for the NC Hamiltonian vector field  $X_{\mathcal{H}_{nc}}.$
	\subsection{ NC Hamilton-Jacobi equation and generating function}
	The NC Hamilton-Jacobi equation is a nonlinear equation given by
	\begin{equation}\label{Eq_3_17}
	E_{nc} = \mathcal{H}_{nc}\Bigg(q,\dfrac{\partial{W^{nc}}}{\partial{q}}\Bigg).
	\end{equation}
	Thus,
	\begin{equation*}
	E_{nc}  = \dfrac{1}{2}\Bigg\{-\Bigg(\dfrac{\partial{W^{nc}}}{\partial{q_{1}}}\Bigg)^{2} +  \sum_{k=2}^{4}\Bigg(\dfrac{\partial{W^{nc}}}{\partial{q_{k}}} + \dfrac{1}{2}\sum_{j=2}^{4}\lambda_{kj}q_{j} \Bigg)^{2} \Bigg\},
	\end{equation*}
	where $E_{nc}$ is a constant.
	Setting  $W^{nc} = \displaystyle\sum_{i=1}^{4}W^{nc}_{i}(q_{i}),$  where $W^{nc}_{i}(q_{i})=a_{i}q_{i}$ and $a_{i}$ $(i=1,2,3,4)$ are constants, not depending on   $q_{i},$
	leads to  $a_{i}= \dfrac{\partial{W^{nc}_{i}}}{\partial{q_{i}}},$ and  \eqref{Eq_3_17} becomes
	\[
	2E_{nc} = - a_{1}^{2} + \sum_{k=2}^{4}\Bigg(a_{k} + \dfrac{1}{2}\sum_{j=2}^{4}\lambda_{kj}q_{j} \Bigg)^{2}.
	\]
	Assume now $\bigg[\displaystyle\sum_{k=2}^{4}\Bigg(a_{k} + \dfrac{1}{2}\displaystyle\sum_{j=2}^{4}\lambda_{kj}q_{j} \Bigg)^{2} - 2E_{nc}\bigg] > 0.$ Then, 
	
	\begin{equation*} \label{Eq_3_18}
	a_{1} =  \pm\sqrt{\sum_{k=2}^{4}\Bigg(a_{k} + \dfrac{1}{2}\sum_{j=2}^{4}\lambda_{kj}q_{j} \Bigg)^{2} - 2E_{nc}}.
	\end{equation*}
	Considering the future domain yields
	\begin{equation} \label{Eq_3_18}
	a_{1} =  \sqrt{\sum_{k=2}^{4}\Bigg(a_{k} + \dfrac{1}{2}\sum_{j=2}^{4}\lambda_{kj}q_{j} \Bigg)^{2} - 2E_{nc}},
	\end{equation}
	and 
	\begin{equation} \label{Eq_3_19}
	W^{nc} = \Bigg( \sqrt{\sum_{k=2}^{4}\Bigg(a_{k} + \dfrac{1}{2}\sum_{j=2}^{4}\lambda_{kj}q_{j} \Bigg)^{2} - 2E_{nc}} \Bigg)q_{1} +  \sum_{k=2}^{4}a_{k}q_{k}.
	\end{equation}
	Now, we introduce a generating function by using the above solution such that
	$W^{nc} = W^{nc}(q_{1},q_{2},q_{3},q_{4},Q_{1},Q_{2},Q_{3},Q_{4})$ becomes
	\begin{equation} \label{Eq_3_20}
	W^{nc} = \Bigg(\sqrt{\sum_{k=2}^{4}Q^{2}_{k} - 2Q_{1}} \Bigg)q_{1} + \sum_{k=2}^{4}\Bigg(Q_{k} - \dfrac{1}{2}\sum_{j=2}^{4}\lambda_{kj}q_{j}\Bigg) q_{k},
	\end{equation}
	where
	\begin{equation}
	\bigg(\displaystyle\sum_{k=2}^{4}Q^{2}_{k} - 2Q_{1}\bigg) > 0, \ Q_{1}=E_{nc},
	\end{equation} and
	\begin{equation}
	Q_{k} = a_{k} + \dfrac{1}{2}\displaystyle\sum_{j=2}^{4}\lambda_{kj}q_{j},  (k=2,3,4).\end{equation} Thanks to the canonical transformations \eqref{can}, 
	we obtain the following relationship between the canonical coordinate system $(P,Q)$ and the original coordinate system $(p,q):$
	\begin{equation}\label{Eq_3_21}
	\left\{
	\begin{array}{ll}
	p_{1}= \sqrt{\displaystyle\sum_{k=2}^{4}Q^{2}_{k} - 2Q_{1}}   \\
	p_{k}=  Q_{k}   \\
	q_{1}=P_{1}\sqrt{\displaystyle\sum_{k=2}^{4}Q^{2}_{k} - 2Q_{1}}   \\
	q_{k}= -P_{k} - Q_{k}P_{1}
	\end{array}
	\right. ; \quad \left\{
	\begin{array}{ll}
	P_{1}=\dfrac{q_{1}}{p_{1}}   \\
	P_{k}=-\dfrac{p_{k}q_{1}}{p_{1}} - q_{k}  \\
	Q_{1} = \mathcal{H}_{nc} \\
	Q_{k}= p_{k},
	\end{array}
	\right.
	\end{equation}
	where $k=2,3,4.$
	\subsection{(1,1)-tensor field T as recursion operator}
	In the $(P,Q)$ coordinate systems, the Hamiltonian vector field is defined by
	\[
	X_{\mathcal{H}_{nc}} := \{\mathcal{H}_{nc},.\}_{nc} = \sum_{\nu=1}^{4}\theta_{\nu}^{-1}\Bigg(\dfrac{\partial{\mathcal{H}_{nc}}}{\partial{P_{\nu}}}\dfrac{\partial}{\partial{Q_{\nu}}} - \dfrac{\partial{\mathcal{H}_{nc}}}{\partial{Q_{\nu}}}\dfrac{\partial}{\partial{P_{\nu}}}\Bigg).
	\]
	Setting $\mathcal{H}_{nc} = Q_{1}$ and $\theta_{1} = 1$ transforms  the  NC Hamiltonian vector field $X_{\mathcal{H}_{nc}}$ and symplectic form $\omega_{nc}$ into the forms
	\begin{equation} \label{Eq_3_22}
	X_{\mathcal{H}_{nc}}= - \dfrac{\partial{\mathcal{H}_{nc}}}{\partial{Q_{1}}}\dfrac{\partial}{\partial{P_{1}}}= -\dfrac{\partial}{\partial{P_{1}}},
	\end{equation}
	and
	\begin{equation} \label{Eq_3_23}
	\omega_{nc}= \sum_{\nu=1}^{4}\theta_{\nu}dP_{\nu}\wedge dQ_{\nu},
	\end{equation}
	respectively.
	A tensor field $T$ of $(1,1)$-type can then be expressed as :
	\begin{equation} \label{Eq_3_23}
	T = \sum_{\nu=1}^{4}Q_{\nu}\Bigg(\dfrac{\partial}{\partial{P_{\nu}}}\otimes dP_{\nu} + \dfrac{\partial}{\partial{Q_{\nu}}}\otimes dQ_{\nu}\Bigg).
	\end{equation}
	Letting $x_{\nu}=Q_{\nu}$ and $x_{\nu +4} = P_{\nu}$, where $\nu = 1,2,3,4,$ affords 
	the tensor field  
	\[
	T
	= \sum_{i,j =1}^{8}T^{i}_{j}\dfrac{\partial}{\partial{x^{i}}}\otimes dx^{j},
	\]
	with $x\equiv (Q_{1},...,Q_{4}, P_{1},...,P_{4}).$ The matrix $(T^{i}_{j})$ is given by
	\[
	(T^{i}_{j}) = \left(
	\begin{array}{cc}
	^{t}A & O \\
	O & A \\
	\end{array}
	\right), \ \ \
	(A_{j}^{i}) =\left(
	\begin{array}{cccc}
	Q_{1} & 0 & 0 & 0 \\
	0 & Q_{2} & 0 & 0 \\
	0 & 0 & Q_{3} & 0 \\
	0 & 0 & 0 & Q_{4} \\
	\end{array}
	\right).
	\]
	Then, by  Lemma \ref{lem_1}, $T$ satisfies $\mathcal{L}_{X_{\mathcal{H}_{nc}}}T = 0$, the Nijenhuis torsion $ \mathcal{N}_{T}$ of $T$ is vanishing, i.e.  $ \mathcal{N}_{T} = 0,$   and $deg Q_{i} = 2$. Hence, $T$ is a recursion operator of the Hamiltonian vector field $X_{\mathcal{H}_{nc}}$. The constants of motion $Tr(T^{l}), \ (l = 1,2,3,4),$ of the geodesic flow
	are :
	\[
	Tr(T^{l}) = 2(Q_{1}^{l} + Q_{2}^{l} + Q_{3}^{l} +Q_{4}^{l}), \quad l = 1,2,3,4.
	\]
	
	\begin{proposition}
		Assume:
		\begin{enumerate}
			\item [	(1)]   	$\lambda_{1\mu} = \alpha_{1\mu} = 0, \ \ \mu= 1, \ 2, \ 3, \ 4 ;$
			\item [	(2)] 	$\lambda_{\nu\mu}\theta_{\nu} = \lambda_{\mu\nu}\theta_{\mu}, \quad \mbox{for every} \ \nu,\mu =2,\ 3,\ 4.$
		\end{enumerate}
		Then, the geodesic flow  has a recursion operator $T$ given by
		\begin{equation} \label{Eq_3_24}
		T = \sum_{\mu,\nu =1}^{4}\Bigg(M^{\mu}_{\nu}\dfrac{\partial}{\partial{q_{\nu}}}\otimes dq_{\mu} + N^{\mu}_{\nu}\dfrac{\partial}{\partial{p_{\nu}}}\otimes dp_{\mu} + L^{\mu}_{\nu}\dfrac{\partial}{\partial{q_{\nu}}}\otimes dp_{\mu} +R^{\mu}_{\nu}\dfrac{\partial}{\partial{p_{\nu}}}\otimes dq_{\mu}\Bigg),
		\end{equation}
		where
		
		{\scriptsize
			\[
			M = \left(
			\begin{array}{cccc}
			\mathcal{H}_{nc}&\dfrac{p_{2}}{p_{1}}\bigg( p_{2} - \mathcal{H}_{nc}\bigg) &  \dfrac{p_{3}}{p_{1}}\bigg( p_{3} - \mathcal{H}_{nc}\bigg)& \dfrac{p_{4}}{p_{1}}\bigg( p_{4} - \mathcal{H}_{nc}\bigg) \\
			\\
			\dfrac{q_{1}\mathcal{H}_{nc}}{p_{1}^{2}}S_{2} & p_{2} & 0& 0 \\
			\\
			\dfrac{q_{1}\mathcal{H}_{nc}}{p_{1}^{2}}S_{3}    & 0 & p_{3} & 0 \\
			\\
			\dfrac{q_{1}\mathcal{H}_{nc}}{p_{1}^{2}}S_{4} & 0 & 0 &  p_{4}
			\end{array}
			\right),
			\]}
		
		{\scriptsize
			\[
			N =\left(
			\begin{array}{cccc}
			\mathcal{H}_{nc} & 0 & 0 & 0 \\
			\\
			\dfrac{p_{2}}{p_{1}}V_{2}& p_{2}  & 0 & 0 \\
			\\
			\dfrac{p_{3}}{p_{1}}V_{3} & 0 & p_{3} & 0 \\
			\\
			\dfrac{p_{4}}{p_{1}}V_{4} & 0 & 0 & p_{4}
			\end{array}
			\right),
			\]}
		
		{\scriptsize
			\[
			L =\left(
			\begin{array}{cccc}
			0 & \dfrac{q_{1}p_{2}}{p_{1}^{2}}\bigg[\dfrac{p_{2}}{p_{1}}\bigg(\mathcal{H}_{nc} - p_{2} \bigg)\bigg] & \dfrac{q_{1}}{p_{1}}\bigg[\dfrac{p_{3}}{p_{1}}\bigg(\mathcal{H}_{nc} - p_{3} \bigg)\bigg] & \dfrac{q_{1}}{p_{1}}\bigg[\dfrac{p_{4}}{p_{1}}\bigg(\mathcal{H}_{nc} - p_{4} \bigg)\bigg]\\
			\\
			\dfrac{q_{1}p_{2}}{p_{1}^{2}}V_{2} & 0& 0 & 0 \\
			\\
			\dfrac{q_{1}p_{2}}{p_{1}^{2}}V_{3} & 0 & 0 & 0 \\
			\\
			\dfrac{q_{1}p_{2}}{p_{1}^{2}}V_{4} & 0 & 0 & 0 \\
			\end{array}
			\right),
			\]}
		
		{\scriptsize
			\[
			R =\left(
			\begin{array}{cccc}
			0& 0& 0& 0\\
			\\
			\dfrac{\mathcal{H}_{nc}}{p_{1}}S_{2}& 0& 0 & 0 \\
			\\
			\dfrac{\mathcal{H}_{nc}}{p_{1}}S_{3}& 0 & 0 & 0 \\
			\\
			\dfrac{\mathcal{H}_{nc}}{p_{1}}S_{4}& 0 & 0 & 0 \\
			\end{array}
			\right),
			\]}
		with
		$V_{k} = p_{k} - \mathcal{H}_{nc} - \dfrac{\mathcal{H}_{nc}}{2p_{k}}\displaystyle\sum_{j=2}^{4}\lambda_{kj}q_{j}$ and $S_{k} = -\dfrac{1}{2} \displaystyle\sum_{i=2}^{^4}\lambda_{ik}\Omega_{i},  (k= 2,\ 3,\ 4).$ \\
		The constants of motion in the original coordinate system $(p,q)$ are $Tr(T^{l}),\\ (l =1,\ 2, \ 3,\ 4) :$
		\begin{equation}
		Tr(T^{l})  = 2\mathcal{H}_{nc}^{l} + 2(p_{2}^{l} + p_{3}^{l} + p_{4}^{l})  = \dfrac{1}{2^{l - 1}}(-p_{1}^{2} + \sum_{k=2}^{4} \varpi_{k}^{2})^{l} + 2(p_{2}^{l}+ p_{3}^{l} + p_{4}^{l}).
		\end{equation}
	\end{proposition}
	
	\section{ Concluding remarks} \label{sec_4}
	In this paper, we have constructed a recursion operator of a Hamiltonian vector field for a geodesic flow in a noncommutative Minkowski phase space, and computed the  associated constants of motion. For the vanishing deformation parameter $\beta,$ the NC Minkowski phase space turns to be the usual  one, and all the results displayed in this work are reduced to the particular cases examined in \cite{t} and \cite{ta}.

\end{document}